\documentclass{osa-article}

\journal{oe}

\begin{document}

\title{Ghost imaging LiDAR via sparsity constraints using push-broom scanning}
\author{Shuang Ma,\authormark{1,2} Zhentao Liu,\authormark{1,*} Chenglong Wang,\authormark{1,2} Chenyu Hu,\authormark{1,2} Enrong Li,\authormark{1} Wenlin Gong,\authormark{1} Zhishen Tong,\authormark{1,2} Jianrong Wu,\authormark{1} Xia Shen,\authormark{1} and Shensheng Han\authormark{1}}

\address{\authormark{1}Key Laboratory for Quantum Optics and Center for Cold Atom Physics of CAS, Shanghai Institute of Optics and Fine Mechanics, Chinese Academy of Sciences, Shanghai, 201800, China;\\
\authormark{2}Center of Materials Science and Optoelectronics Engineering, University of Chinese Academy of Sciences, Beijing 100049, China\\}

\email{\authormark{*}ztliu@siom.ac.cn} 



\begin{abstract}
{Ghost imaging LiDAR via sparsity constraints using push-broom scanning is proposed. It can image the stationary target scene continuously along the scanning direction by taking advantage of the relative movement between the platform and the target scene. Compared to conventional ghost imaging LiDAR that requires multiple speckle patterns staring the target, ghost imaging LiDAR via sparsity constraints using push-broom scanning not only simplifies the imaging system, but also reduces the sampling number. Numerical simulations and experiments have demonstrated its efficiency.}
\end{abstract}

\section{Introduction}
LiDAR \cite{elachi1988spaceborne} is a novel detection tool that has been widely applied in remote sensing. Most of the commercial LiDAR systems are whiskbroom LiDAR, which obtain the target information by the optical-mechanical scanning mechanism. It exploits a whiskbroom mirror to scan in one dimension across the aircraft flight path, and the physical motion of the aircraft itself to scan along the flight path. It is very sensitive and low-cost, but heavily suffers from optical-mechanical pointing errors and bad real-time imaging of moving targets \cite{national2014laser}. Another type of LiDAR, as known as push-broom LiDAR, uses a line array detector across the aircraft flight path instead of the optical-mechanical scanning. The elimination of the optical-mechanical scanning greatly improves the system stabilization and the real-time  performance. However, it is less sensitive than the whiskbroom LiDAR and more easily influenced by the background noise \cite{national2014laser}. However, in both LiDARs, the target information is acquired by point-to-point corresponding method between the target and the detector plane, therefore, the correlation between pixels of the target is not involved in the imaging process. In addition, all these imaging methods require sampling rate at or beyond the Nyquist limit. The redundancy sampling method of both LiDARs greatly limits the further development of the LiDAR techniques, especially for high-resolution imaging in long ranges.

As a novel time-resolved single-pixel LiDAR method, ghost imaging LiDAR via sparsity constraint (GISC LiDAR) \cite{sun2012normalized,shapiro2012physics,shapiro2008computational,smith2018turbulence,shih2012physics,ferri2010differential,cheng2004incoherent,zhang2005correlated,han2018review,gong2016three,liu2016spectral,yu2016fourier,liu2018lensless,zhao2012ghost,chen2013ghost}, which takes advantages of the sparsity property of nature targets and compressive sensing approach \cite{candes2008introduction,donoho2006compressed, stern2007single,katz2009compressive}, makes the compressive sampling possible \cite{zerom2011entangled,gong2012experimental,gong2012multiple,gong2013experimental}. On the basis of its unique information acquisition mechanism, GISC LiDAR realizes high resolution imaging and keeps the high sensitivity and low-cost merits, and it has the potential of real-time imaging the moving targets \cite{li2014ghost,zhang2013improving,li2015ghost,yu2014single,yu2016compressive,pan2017moving,yu2015compressive}. However, as a typical staring imaging mechanism, traditional single-pixel GISC LiDAR always requires a great number of measurements for practical remote sensing application.  Because of the relative movement caused by aircraft flight, there still exists the motion blur problem \cite{wang2018airborne}. One successful strategy applied in the reported airborne GISC LiDAR system \cite{wang2018airborne} is the motion compensation technique based on opto-electronic tracking technique. However, this motion compensation mechanism limits its imaging field of view centered at the focusing area of the motion compensation system, which means a continuous remote sensing will be hard to be realized. And the opto-electronic tracking method might be disabled in the cases such as at night or low visibility weather. Another possible scheme to solve this motion blur problem is referring to the traditional push-broom mechanism. In this letter, we propose ghost imaging LiDAR via sparsity constraints using push-broom scanning (push-broom GISC LiDAR). In this method, the target scene can be imaged continuously along the scanning direction by taking advantage of the relative movement between the platform and the target scene. The size of the target is no longer limited by the field of view, and the motion compensation is not required.

\section{The model and methods}

\begin{figure}[ht]
\centering
\includegraphics[width=12 cm]{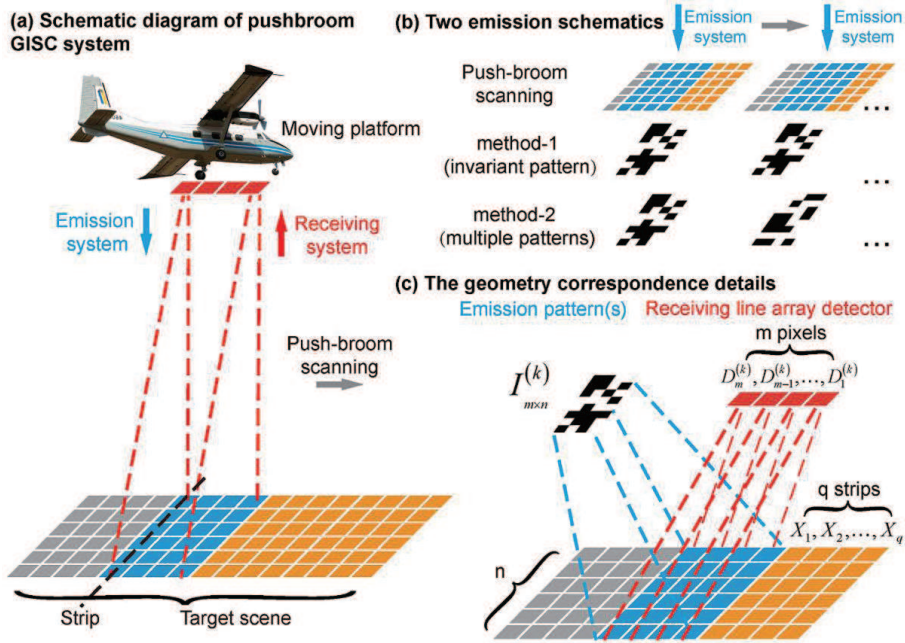}
\caption{Schematic of the push-broom GISC LiDAR system. (a) Schematic diagram of push-broom GISC LiDAR system. (b) Two emission schematics. (c) The geometry correspondence details.}
\label{Fig:1} 
\end{figure}

Figure \ref{Fig:1} presents the schematic of the proposed push-broom GISC LiDAR. The push-broom GISC LiDAR system, mounted on a moving platform, emits predetermined and calibrated rectangular speckle pattern(s) to cover a certain region of the target scene (The blue area in Fig. \ref{Fig:1}(a)) in each laser pulse. Then a receiving system focuses the rectangular receiving speckle pattern(s) to a line array detector through an orthogonal cylindrical lens group. Across the flight direction, the imaging region is divided into many strips. The signal of each pixel of the line array detector corresponds to a distinct strip of this target region; and as the system moves along the flight path strip by strip (push-broom scanning), every strip can then be scanned and sampled multiple times sequentially by each pixel of the linear detector. If the sampling frequency of the emission system and the receiving system $f$ can be well synchronized with the push-broom velocity $v$ across each strip ($v=rf$, where $r$ denotes the resolution of the strip along the push-broom direction.), a continuous sensing along the push-boom direction can be realized.

The system may either use an invariant illuminating pattern or multiple illuminating patterns during the sampling process, denoted as method-1 and method-2, respectively (Fig. \ref{Fig:1}(b)). For method-1, the emitting illuminating pattern always keeps invariant during the whole push-broom scanning process (Fig. \ref{Fig:1}(b)). On the contrary, for method-2, the emitting illuminating pattern changes every time when a new strip comes into the focusing region along the flight direction (Fig. \ref{Fig:1}(b)). Nonetheless, these two emission schemes can be described in a general manner as below. For the sake of simplicity, we assume that during the sampling process the platform moves at a fixed direction parallel to the line array detector. Suppose the transverse resolution across the flight direction is $n$ and the pixels of the line array detector in the receiving system is $m$, the sampling rate of this push-broom GISC LiDAR system can be defined as $\eta = m/n$. If we want to realize sensing for $q$ strips along the flight direction, a push-broom scanning for $q+m-1$ strips is required so that each of these scanned strips could be equally sampled $m$ times in this process(Fig. \ref{Fig:1}(c)). Then, we use a $n$-by-1 column vector $X_i$ to represent the $i$-th strip of the target, and denote the $k$-th $m \times n $ illuminated pattern as ${I^{\left( k \right)}}$, and the $k$-th detection signals of the line array detector as ${D^{\left( k \right)}}
$ (a $m$-by-1 vector). With the geometrical relationship of the system, one may write down the equation that relates the detection signal $Y_i$ and $X_i$ with a general form:

\begin{equation}
{Y_i} = {A_i}{X_i}.
\end{equation} 
Here, ${Y_i} = {\left[ {\begin{array}{*{20}{c}}
{D_1^{\left( i \right)}}& \ldots &{D_j^{\left( {i + j - 1} \right)}}& \ldots &{D_m^{\left( {i + m - 1} \right)}}
\end{array}} \right]^T}$ is a $m$-by-1 column vector, where $D_{j}^{(i+j-1)}$ is the $j$-th element of the $D^{(i+j-1)}$. $A_i$ is a $m$-by-$n$ measurement matrix, where the vector form can be represented as

\begin{equation}
A_i = \left[ {\begin{array}{*{20}{c}}
{{I_{1}^{(i)}}}\\
 \vdots \\
{{I_{j}^{(i+j-1)}}}\\
 \vdots \\
{{I_{m}^{(i+m-1)}}}
\end{array}} \right],
\end{equation} 
with $I_{j}^{(i+j-1)}$ ($1 \le j \le m$), a 1-by-$n$ row vector, denoting the $j$-th row of the $(i+j-1)$-th illuminating pattern $I^{(i+j-1)}$.

Analog to the above analyses, after the push-broom scanning for $q + m -1$ strips ($q \ge 1$), one may have the general equation for the total detection signal $Y$ and the total information of these $q$ strips $X$ as follows,

\begin{equation}
Y = AX.
\label{equ:(3)}
\end{equation} 
And similarly the vector form of $Y$ and $X$ can be respectively represented as

\begin{equation}
{Y} = \left[ {\begin{array}{*{20}{c}}
{{Y_1}}\\
 \vdots \\
{{Y_i}}\\
 \vdots \\
{{Y_q}}
\end{array}} \right],
\end{equation} 
and
\begin{equation}
{X} = \left[ {\begin{array}{*{20}{c}}
{{X_1}}\\
 \vdots \\
{{X_i}}\\
 \vdots \\
{{X_q}}
\end{array}} \right].
\end{equation} 
And the total measurement matrix $A$ is the combination of ${A_i}$:

\begin{equation}
{A} = \left[ {\begin{array}{*{20}{c}}
{{{{A}}_1}}&{}&{}&{}&{}\\
{}& \ddots &{}&{}&{}\\
{}&{}&{{{{A}}_i}}&{}&{}\\
{}&{}&{}& \ddots &{}\\
{}&{}&{}&{}&{{{{A}}_q}}
\end{array}} \right].
\end{equation} 

Particularly, for method-1, the emitting illuminating pattern always keeps invariant during the whole push-broom scanning process so that the measurement matrix of every strip is actually the same as each other. Therefore, the sensing formulation for method-1 can be simplified as,

\begin{equation}
{Y}' = A'X',
\label{equ:(7)}
\end{equation} 
where ${Y}' = \left[ {\begin{array}{*{20}{c}}{{Y_1}}& \cdots &{{Y_i}}& \cdots &{{Y_q}}
\end{array}} \right]$ is a $m$-by-$q$ matrix, ${X}' = \left[ {\begin{array}{*{20}{c}}{{X_1}}& \cdots &{{X_i}}& \cdots &{{X_q}}
\end{array}} \right]$ is a $n$-by-$q$ matrix corresponding to the target scene, and $A'$ is the invariant illuminating pattern used in the method-1 push-broom GISC LiDAR.

To reconstruct $X$ from $Y$ according to Eq. (\ref{equ:(3)}), exploiting the sparsity property of the target scene image, here we adopt the total variation (TV) regularization and solve the following optimization program \cite{candes2006robust}, namely, 

\begin{equation}
{X^ * } = \arg \mathop {\min }\limits_X \left\| Y- AX \right\|_2^2 + \lambda {\left\| {TV\left( X \right)} \right\|_1},
\end{equation}

\begin{equation}
{\left\| {TV\left( X \right)} \right\|_1} = \sum\limits_{1 \le i \le qn - n,i \ne kn,k \in N } {\sqrt {{{\left( {{x_{i + 1}} - {x_i}} \right)}^2} + {{\left( {{x_{i + n}} - {x_i}} \right)}^2}} }, 
\end{equation}
where ${x_{i}}$ denotes the $\left( {i} \right)$-th element of the column vector $X$, $N$ stands for natural number.

To reconstruct $X'$ from $Y'$ according to Eq. (\ref{equ:(7)}) (simplification version for method-1), the TV regularization program can be represented as, 

\begin{equation}
{X^ * } = \arg \mathop {\min }\limits_X \left\| Y'- A'X' \right\|_2^2 + \lambda {\left\| {TV\left( X' \right)} \right\|_1},
\end{equation}
\begin{equation}
{\left\| {TV\left( X' \right)} \right\|_1} = \sum\limits_{1 \le i \le n-1,1 \le j \le q - 1} {\sqrt {{{\left( {{x'_{i + 1,j}} - {x'_{i,j}}} \right)}^2} + {{\left( {{x'_{i,j + 1}} - {x'_{i,j}}} \right)}^2}} }, 
\end{equation}
with ${x'_{i,j}}$ the $\left( {i,j} \right)$-th element of the matrix $X'$.

\section{Simulation and experimental results}

\begin{figure}[ht]
\centering
\includegraphics[width=11 cm]{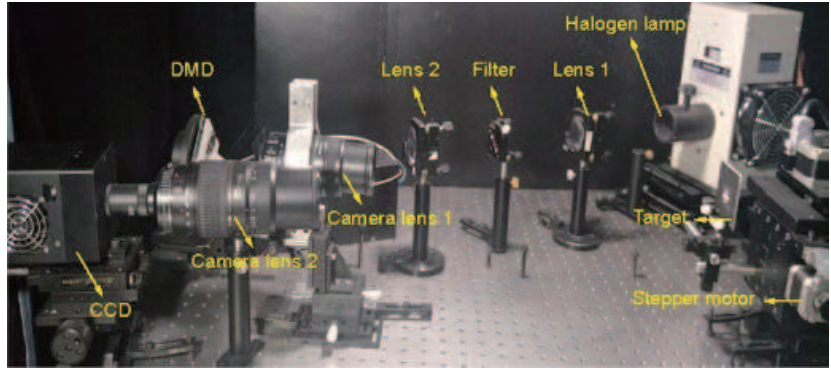}
\caption{Experimental setup of push-broom GISC system. A halogen lamp emitted the light uniformly illuminated onto the DMD, then the speckle pattern reflected by the DMD was imaged onto the target scene, the reflected light of the speckle pattern projected onto target was imaged on CCD.}
\label{Fig:2} 
\end{figure}

To verify the proposed imaging methods, numerical simulations and experiments were performed correspondingly. In the proposed push-broom scanning methods demonstrated in Fig. \ref{Fig:1}, the push-broom GISC LiDAR system was located on a moving platform and the target scene keeps static. In our experiment, a moving target strategy was substituted for convenience, instead of a moving platform to simulate the continuous relative movement process. The experimental setup was illustrated in Fig. \ref{Fig:2}. A halogen lamp emited the light through a Kohler lighting system made up of lens1 and lens2, uniformly illuminating the digital mirror array device (DMD) (XD-ED01(N)). The focal lengths of lens1 and lens2 were ${10cm}$ and ${15cm}$, respectively, and a $10nm$ bandwidth filter centered at $655nm$ was used here to improve the system's image quality. The coded patterns were prefabricated by modulating the mirrors of the DMD. This DMD has $1024 \times 768$ pixels, of which the size was $13.68\mu m \times 13.68\mu m$. We choose $256 \times 256$ pixels from DMD for the experiments with $4 \times 4$ points merged in the pattern element. Each pattern element took the value of either 0 or 1. Then the speckle pattern reflected by the DMD was imaged onto the target scene by camera lens1 (Canon EFS55-250mm F/4-5.6 IS II). The region size of target scene was ${\rm{1}}{\rm{.59cm}} \times {\rm{1}}{\rm{.59cm}}$ and the resolution of each strip is $r=248um$. The reflected light of the pattern projected onto the target was imaged onto CCD by camera lens2 (Tamron AF70-300mm F/4-5.6). Test target scenes were images printed on paper, characterized by some strips. The CCD energy in the vertical direction of motion was integrated working as a line array detector. Because the sampling frequency was set as  $2$Hz in the experiment, the velocity of stepper motor was $v = 496um/s$ base on the synchronization requirement.
  
\begin{figure}[ht]
\centering
\includegraphics[width=12 cm]{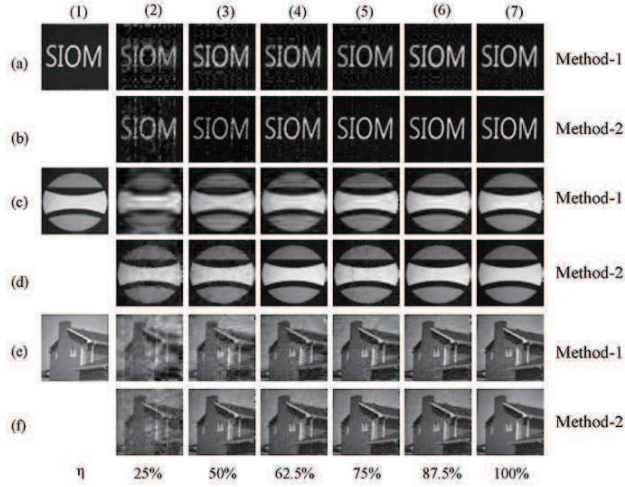}
\caption{Simulated results of method-1 and method-2. (a1) Binary target alphabet "SIOM", (c1) Gray scale target logo pattern, (e1) Gray scale target "house". Columns (2)-(7) correspond to sampling rate $\eta$ are 0.25, 0.5, 0.625, 0.75, 0.875 and 1, respectively. (a2)-(a7), (c2)-(c7) and (e2)-(e7) are the simulated results of method-1. (b2)-(b7), (d2)-(d7) and (f2)-(f7) are the simulated results of method-2.}
\label{Fig:3} 
\end{figure}

\begin{figure}[ht]
\centering
\includegraphics[width=12 cm]{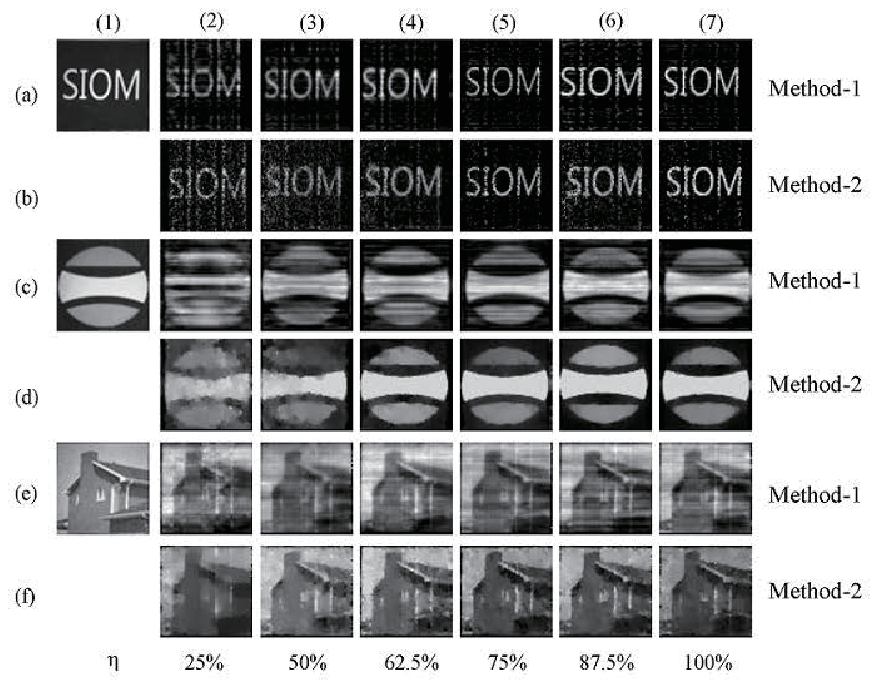}
\caption{Experimental results of method-1 and method-2. (a1) The printed photo of binary target alphabet "SIOM", (c1) the printed photo of the gray scale target logo pattern, (e1) the printed photo of the gray scale target "house". Columns (2)-(7) correspond to sampling rate $\eta$ are 0.25, 0.5, 0.625, 0.75, 0.875 and 1, respectively. (a2)-(a7), (c2)-(c7) and (e2)-(e7) are the experimental results of method-1. (b2)-(b7), (d2)-(d7) and (f2)-(f7) are the experimental results of method-2.}
\label{Fig:4} 
\end{figure}

Firstly, the performance of the compressive sampling capability of the proposed push-broom GISC LiDAR is investigated both simulatively (Fig. \ref{Fig:3}) and experimentally (Fig. \ref{Fig:4}). In both of the Figs. \ref{Fig:3} and Figs. \ref{Fig:4}, the images of column (1) are three different typical targets, from binary target to grayscale target, and simple to complex. And the results shown in columns (2)-(7) correspond to the sampling rate $\eta$ of 0.25, 0.5, 0.625, 0.75, 0.875 and 1, respectively. Rows (a), (c), and (e) are for method-1 and rows (b), (d), and (f) are for method-2. 


\begin{figure}[ht]
\centering
\includegraphics[width=12 cm]{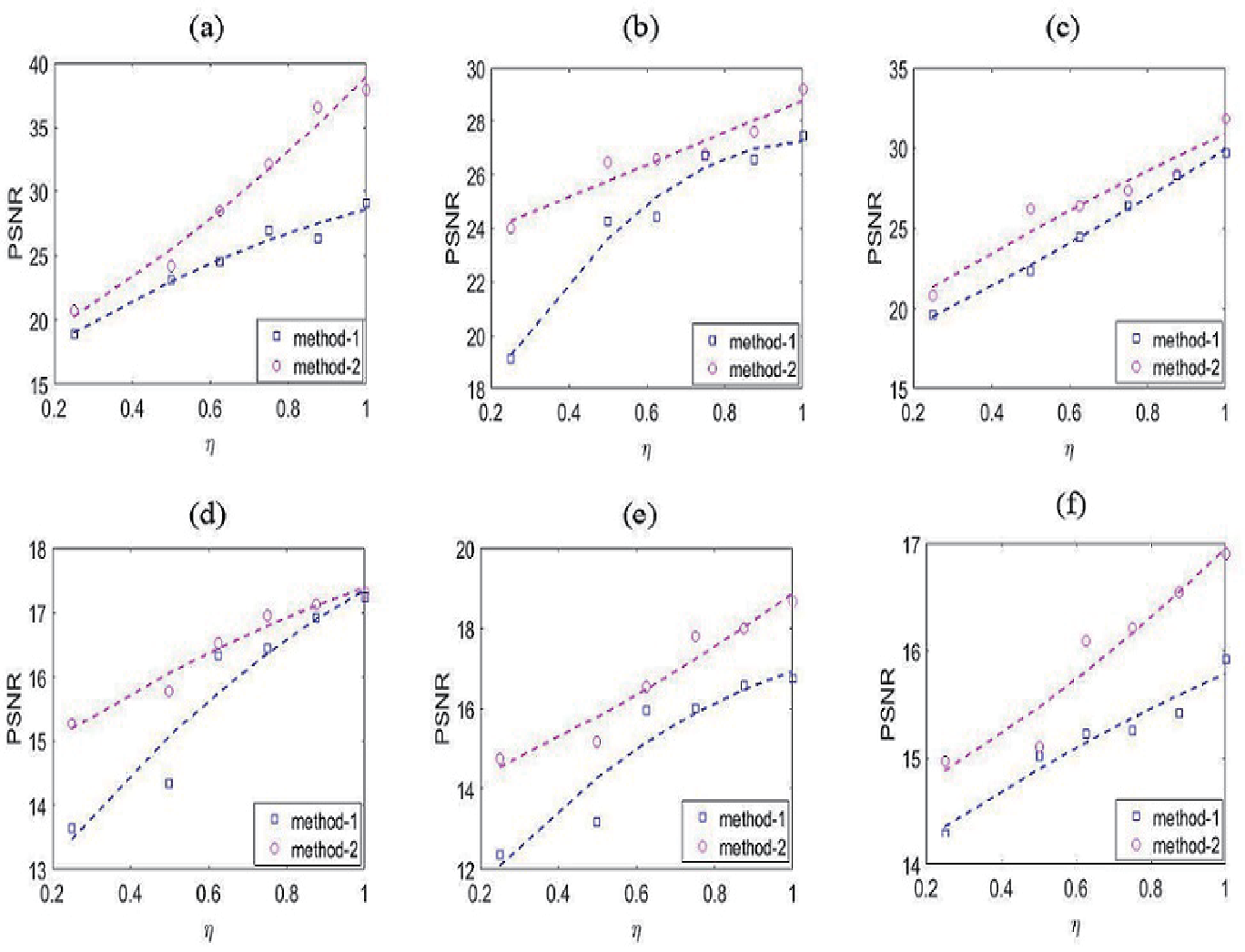}
\caption{PSNR curves with respect to sampling rate $\eta$. The blue and mauve dashed curves signify the fitted PSNR of the target scene reconstructed by method-1 and method-2, respectively. The scattered points of corresponding colors are the PSNR before fitting. (a) PSNR curves of simulated results of the alphabet "SIOM". (b) PSNR curves of simulated results of the logo pattern. (c) PSNR curves of simulated results of the "house". (d) PSNR curves of experimental results of the alphabet "SIOM". (e) PSNR curves of experimental results of the logo pattern. (f) PSNR curves of experimental results of the "house".}
\label{Fig:5} 
\end{figure}

Here, we use the peak signal to noise ratio (PSNR) to evaluate the quality of reconstructed images quantitatively. It is the relative logarithm value that the mean square error $ {\left( {{{\rm{2}}^n} - 1} \right)^2}$ between original image and reconstructed image, the unit of PSNR is dB, namely,
\begin{equation}
PSNR = 10 \times {\log _{10}}\left( {\frac{{{{\left( {{2^n} - 1} \right)}^2}}}{{MSE}}} \right),
\end{equation}
where the MSE is the mean square error between original image and reconstructed image. Figs. \ref{Fig:5}(a), \ref{Fig:5}(b) and \ref{Fig:5}(c) are the PSNR of the simulated results of the target scene in Figs. \ref{Fig:3}(a1), \ref{Fig:3}(c1) and \ref{Fig:3}(e1), respectively. Figs. \ref{Fig:5} (d), \ref{Fig:5}(e) and \ref{Fig:5}(f) are the PSNR of the experimental results of the target scene in Figs. \ref{Fig:3}(a1), \ref{Fig:3}(c1) and \ref{Fig:3}(e1), respectively. The blue and mauve dashed curves signify the fitted PSNR of the target scene reconstructed by method-1 and method-2, respectively. The scattered points of corresponding colors are the PSNR before fitting. The results show that the PSNR of reconstructed image by method-2 is higher than method-1.

\begin{figure}[ht]
\centering
\includegraphics[width=11.2 cm]{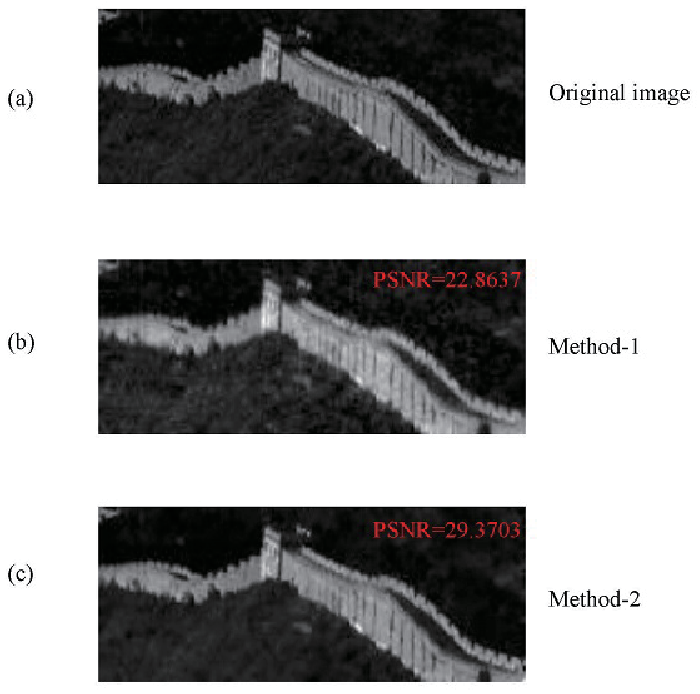}
\caption{The simulated result of continuous the Great Wall image at the $\eta=0.625$. (a) The original image. (b) The simulated result of method-1. (c) The simulated result of method-2.}
\label{Fig:6} 
\end{figure}

After that, a simulation for a continuous scene images with $\eta=0.625$ has been performed to further demonstrate the performance of the push-broom GISC LiDAR. The simulated image is a "Great Wall" picture (Fig. \ref{Fig:6}(a)) with a size of ${\rm{64}} \times {\rm{192}}$ pixels. The simulated results of method-1 and method-2 are shown in Figs. \ref{Fig:6}(b) and \ref{Fig:6}(c), respectively. Note that method-1 takes less time for reconstruction with same computing resources (85.5987 seconds and 96.9540 seconds for method-1 and method-2, respectively, with an Intel(R) Core(TM) i3-2120 CPU @ 3.3 GHz and a 6.00 GB RAM).

\section{Discussion}

As depicted in Fig. \ref{Fig:3}, Fig. \ref{Fig:4} and Fig. \ref{Fig:5}, firstly, all these imaging qualities degrade when the sampling rate decreases, but the reconstructions are still acceptable with compressive sampling. Secondly, according to the three typical types of objects, which represent targets scene from simple to complex, utilized in the simulations and experiments, both methods of the proposed push-broom GISC LiDAR are demonstrated. Thirdly, as presented in Fig. \ref{Fig:3} and Fig. \ref{Fig:4}, the reconstruction with low sampling rate of method-1 schematic have obvious regular noise which seems like there are some transverse strings overlapping on the results. But there are no such obvious phenomenon in the corresponding reconstructions of method-2. This interesting phenomenon can be explained as follows. In method-1, all strips share the same measurement matrix. Therefore, if the targets scene in different strips are very similar with each other, for example, the alphabet "I" and some parts of "M", the noise of the reconstructions are very similar, which brings about those regular visual noise results. On the contrary, in method-2, every strip has its own measurement matrix. Hence even when the target scene is similar, the reconstruction does not accord with each other usually. Therefore, although method-1 requires no dynamic light modulation device, which seems more stable and economic, and the computation cost can be saved, its imaging quality is still need further improved. In addtion, the simulated results demonstrated in Fig. \ref{Fig:6} shows that our proposed push-broom GISC LiDAR can realize remote sensing for a continuous target scene along the scanning direction, which indicates this novel schematic have the potential to develop to be a novel LiDAR technique for remote sensing. However, even though method-2 realizes a better reconstruction, the method-1 is more advantageous than method-2 in the computation time.

For traditional push-broom LiDAR, because the line array detector is across the flight direction, the pixel number of the line array detector must be equal to the resolution number of every strip. In contrast, our proposed push-broom GISC LiDAR can realize the same resolution $n$, with only a $m$ pixels line array detector ($m \le n$) due to its compressive sampling capability. Therefore, push-broom GISC LiDAR could be more economic and more easy to be generalized for higher resolution remote sensing application, than the traditional Push-broom LiDAR. Compared to the previous GISC LiDAR system based on motion compensation technique, target scene now can be imaged continuously along the scanning direction, rather than only focusing on a certain area. And the application of the line array detection mechanism transforms the previous direct two-dimensional reconstruction for a rectangular area to a sequential one-dimensional reconstruction for a strip, which dramatically decreases the required measurements for a single strip. In addition, without the complicated motion compensation system, there is no trouble of the tracking accuracy problem any more, and the performance under low-visible condition can also be further improved.   

Clearly, much optimization for this proposed push-broom GISC LiDAR system still need to be further performed and we might pay close attention on the following aspects in the future work. First of all, for method-2, the emission scheme can be further optimized to reduce the computational scale by repeating the speckles for some period of strips. Secondly, although the method-1 seems more convenient in realization and fast in computation, its imaging quality need further improvements by approaches such as speckle pattern optimization\cite{chen2014application}. What's more, because in the sampling process, both the shake of the moving platform and the variation of the flight speed could heavily influence the final reconstruction, great attentions on how to reduce these impacts will be paid in the future work.

\section{Conclusion}

In conclusion, a push-broom GISC LiDAR is proposed and demonstrated experimentally. Because of its compressive sampling capability, the higher imaging resolution across the push-broom direction can realize by a line array detector with less pixels. With further improvements, the push-broom GISC system might find many useful applications, such as automotive LiDAR, airborne LiDAR, and so on. 

\section*{Funding}
This research was funded by the Hi-Tech Research and Development Program of China in Projects (No.2013AA122902, No.2013AA122901, No.2011AA120102 and No.2011AA120101).\\

\bibliography{sample}

\end{document}